\documentclass[12pt]{article}
\usepackage[top=1in, bottom=1in, left=1in, right=1in]{geometry}
\usepackage[english]{babel}
\usepackage{atbegshi,cite}
\usepackage{amsmath,amssymb,amsbsy,amstext, amsthm, simplewick}
\usepackage{hyperref}
\usepackage{graphicx}
\usepackage{amsfonts}
\usepackage[small]{caption}
\usepackage{upgreek}
\usepackage[titletoc]{appendix}
\usepackage{setspace}
\usepackage{color}

\newcommand{\newc}{\newcommand}
\newc{\fpi}{f_{\pi}}
\newc{\hetap}{\tilde\eta^{\prime}}
\newc{\hpi}{\tilde\pi}
\newc{\etap}{\eta^{\prime}}
\newc{\llll}{\langle\lambda\lambda\rangle}
\newc{\FFd}{F^a\tilde F^a}
\newc{\qbar}{{\overline q}}
\newc{\TR}{{\rm Tr}}
\newc{\Kahler}{K\"ahler }
\newc{\Zbb}{{\mathbb Z}}
\newc{\Rt}{{\mathbb R}^3}
\newc{\Rf}{{\mathbb R}^4}
\newc{\So}{{\mathbb S}^1}
\newc{\zt}{{\mathbb Z}_2}
\newc{\RtSo}{{\mathbb R}^3\times{\mathbb S}^1}
\newc{\scriminus}{{\cal I}^-}
\newc{\scriplus}{{\cal I}^+}
\newc{\mpl}{M_p}
\newc{\Ricci}{\mathcal{R}}
\newc{\bv}{\phi}
\newc{\calU}{{\cal U}}
\newc{\calK}{K}
\newc{\calUi}{{\cal U}^{-1}}
\newc{\calG}{{\cal G}}
\newc{\calO}{{\cal O}}
\newc{\calOb}{{\cal O}^\dagger}
\newc{\hphi}{{\hat\phi}}

\begin{document}
\begin{titlepage}
\begin{flushright}
{\large 
ACFI-T18-02\\
}
\end{flushright}

\vskip 2.0cm

\begin{center}

{\large \bf Domain Walls and the $CP$ Anomaly\\
 in Softly Broken Supersymmetric QCD}

\vskip 1.4cm

{Patrick Draper}
\\
\vskip 1cm
{\it 
Amherst Center for Fundamental Interactions, Department of Physics,\\ University of Massachusetts, Amherst, MA 01003
}\\
\vspace{0.3cm}
\vskip 4pt

\vskip 1.5cm

\begin{abstract}
In ordinary QCD with light, degenerate, fundamental flavors, $CP$ symmetry is spontaneously broken at $\theta=\pi$, and domain wall solutions connecting the vacua can be constructed in chiral perturbation theory. In some cases the breaking of $CP$ saturates an 't Hooft anomaly, and anomaly inflow requires nontrivial massless excitations on the domain walls. Analogously, $CP$ can be spontaneously broken in supersymmetric QCD with light flavors and small soft breaking parameters. We study $CP$ breaking and domain walls in softly broken SQCD with $N_f<N$ flavors. Relative to ordinary QCD, the supersymmetric case contains an extra light field, the $\etap$, which has interesting effects on the structure of the walls.  Vanishing of the $CP$ anomaly is associated with the existence of multiple domain wall trajectories through field space, including walls which support no nontrivial massless excitations. In cases with an anomaly such walls are forbidden, and their absence in the relevant SQCD theories can be seen directly from the geometry of the low energy field space. In the case $N_f=N-1$, multiple approximately-BPS walls connect the vacua. Corrections to their tensions can be computed at leading order in the soft breaking parameters, producing a phase diagram for the stable wall trajectory. We also comment on domain walls in the similar case of QCD with an adjoint and fundamental flavors, and on the impact of adding an axion in this theory.
\end{abstract}

\end{center}

\vskip 1.0 cm

\end{titlepage}
\setcounter{footnote}{0} \setcounter{page}{2}
\setcounter{section}{0} \setcounter{subsection}{0}
\setcounter{subsubsection}{0}
\setcounter{figure}{0}

\onehalfspacing



\section{Introduction}
The $\theta$ dependence of QCD and QCD-like theories 
is an important probe of nonperturbative phenomena. In real QCD, the vacuum energy is a smoothly varying function of $\theta$, calculable in chiral perturbation theory (ChPT) and exhibiting manifest $2\pi$ periodicity.  At large $N$, the $\eta^\prime$ is light and the vacuum energy becomes a multi-branched function of $\theta$~\cite{witten}. At finite $N$ but with a higher degree of degeneracy among the $N_f$ light fundamental flavors, the potential exhibits $N_f$ local minima and the vacuum energy is again multi-branched. At $\theta=\pi$, $CP$ is spontaneously broken in the global minima~\cite{dashen,witten}.  

When $CP$ is spontaneously broken, the $CP$-conjugate vacua can be connected by a domain wall. $CP$  walls in QCD-like theories with light fundamental flavors were  recently studied in~\cite{qcdcpdw}. For $N_f>1$, domain walls follow calculable trajectories through the pion field space of ChPT, spontaneously breaking the vector-like $SU(N_f)$ flavor symmetry. As a result, the walls support Goldstone modes among their three-dimensional massless excitations.

It was also shown in~\cite{qcdcpdw} that for $\gcd(N,N_f)>1$ and $\theta=\pi$, there is an 't Hooft anomaly between $CP$ and a certain discrete flavor symmetry intertwined with the center symmetry. Anomaly inflow~\cite{Callan:1984sa,Faddeev:1985iz} then requires that in these theories, domain walls connecting the $CP$ vacua must support nontrivial massless degrees of freedom beyond the trivial center of mass mode. Similar anomalies and inflow on domain walls were described in pure gauge theory in~\cite{ttt}, and a sample of other recent  work on center anomalies, intertwined center-flavor symmetries, and related topics includes~\cite{Cherman:2017tey,Komargodski:2017smk,Kitano:2017jng,Tanizaki:2017mtm}.

We will study spontaneous $CP$ breaking in supersymmetric QCD with $N_f<N$ light flavors and  soft SUSY-breaking mass terms. $CP$ and its domain walls have several interesting aspects in these theories:

\begin{itemize}

\item The $CP$ anomaly in softly broken SQCD is the same as the ordinary nonsupersymmetric QCD anomaly described in~\cite{qcdcpdw}. This is because the latter is obtained from the former by adding large mass terms for the scalars and increasing the gaugino mass, which does not change the anomalies\footnote{We thank Zohar Komargodski for a helpful discussion of this point.}  (as long as the limit does not change the value of the topological angle associated with the anomaly; this will be discussed more thoroughly in Sec.~\ref{sec:anomalies}.) 

\item Due to mixing with the $U(1)_R$ symmetry, the $``\eta^\prime"$ associated with spontaneous $U(1)_A$ breaking is light in SQCD even at small $N$.\footnote{The $\etap$ in SUSY at large $N$ was discussed in~\cite{Dine:2016sgq}.}  This extra degree of freedom can permit domain walls along new trajectories in field space relative to ordinary QCD. Such $\eta^\prime$ walls do not support any nontrivial massless excitations, and in fact we will see that they arise between $CP$ vacua precisely when $\gcd(N,N_f)=1$, where the anomaly vanishes. SQCD thus reflects the anomaly in a geometric way: its absence  is associated with the existence of $\etap$ $CP$ walls. 

\item In some cases the $CP$ walls are approximately BPS-saturated, with calculable tensions~\cite{dvalishifman}. BPS domain walls in SQCD have been extensively studied in the literature~\cite{dvalishifman,smilga2,smilga3,smilga,kaplunovsky,deCarlos1,deCarlos2,acharyavafa,ritz1,ritz2}.
SUSY-breaking effects on the tension may be computed perturbatively. We will also see cases where $\etap$ $CP$ walls do not exist in the supersymmetric limit (BPS or not), but then ``come in from infinity" in the presence of small soft breaking.

\item In cases where both BPS $\etap$ walls and BPS pion-like walls (breaking the $SU(N_f)_V$ symmetry) exist, their tensions are equal in the SUSY limit. We can compare the $\Delta T$s for the different trajectories in the space of small soft breakings and find a phase diagram for stability and metastability of the different types of walls.

\end{itemize}

This paper is organized as follows. In Sec.~\ref{sec:review} we review the vacuum structure of SQCD, focusing  on periodicities and discrete gauge symmetries on the pseudomoduli space. We plot the $\eta^\prime$ direction together with the gaugino condensate to make clear the monodromies and the connections between different branches. In Sec.~\ref{sec:CP} we add small soft breaking terms and derive the conditions for spontaneous $CP$ breaking in the global minima. In Sec.~\ref{sec:walls} we study domain walls connecting the $CP$ vacua, discussing some of their general properties and constructing numerical solutions. We show that  only pion walls  exist when $\gcd(N,N_f)>1$. For generic $N_f$ with $\gcd(N,N_f)=1$ we find $\eta^\prime$ walls, which depend sensitively on the presence of SUSY-breaking. In $N_f=N-1$, BPS  $\eta^\prime$ walls exist in the SUSY limit. We compute the corrections to their tensions from soft breaking effects and compare to the tension of walls in pion directions. We discuss the various connections to 't Hooft anomalies in Sec.~\ref{sec:anomalies}. In Sec.~\ref{sec:qcdadj} we comment briefly on domain walls in the theory where the scalars are completely decoupled. 

\section{Supersymmetric Theory}
\label{sec:review}
Here we review standard results in supersymmetric QCD with $N_f<N$ flavors, repackaging the low-energy field space and vacua in a way that is convenient for exhibiting domain wall trajectories.

\subsection{Moduli Space}

The effective superpotential is
\begin{align}
W&=W_{ADS}+m\,\TR\,Q\bar Q\;,\nonumber\\
W_{ADS}=(&N-N_f)\,\frac{\Lambda^{\frac{3N-N_f}{N-N_f}}}{\left(\det Q\bar Q\right)^\frac{1}{N-N_f}}\,e^{i\frac{\theta+2\pi k}{N-N_f}}\;.
\end{align}
For simplicity, we have taken the quark mass matrix to be proportional to the identity, preserving the $SU(N_f)_V$ flavor symmetry. $\Lambda$ is the dynamical scale associated with the UV $SU(N)$ theory. Anticipating the addition of soft masses, particularly a soft gaugino mass, we have left $\theta$ explicit instead of rotating it away. The integer $k=0\dots N-N_f-1$ labels the branches of the gaugino bilinear, which is related to the superpotential via
\begin{align}
\langle\lambda\lambda\rangle&=-\frac{32\pi^2}{N-N_f}W_{ADS}\;.
\label{eq:ll}
\end{align}

\begin{figure}[t!]
\begin{center}
\includegraphics[width=0.5\linewidth]{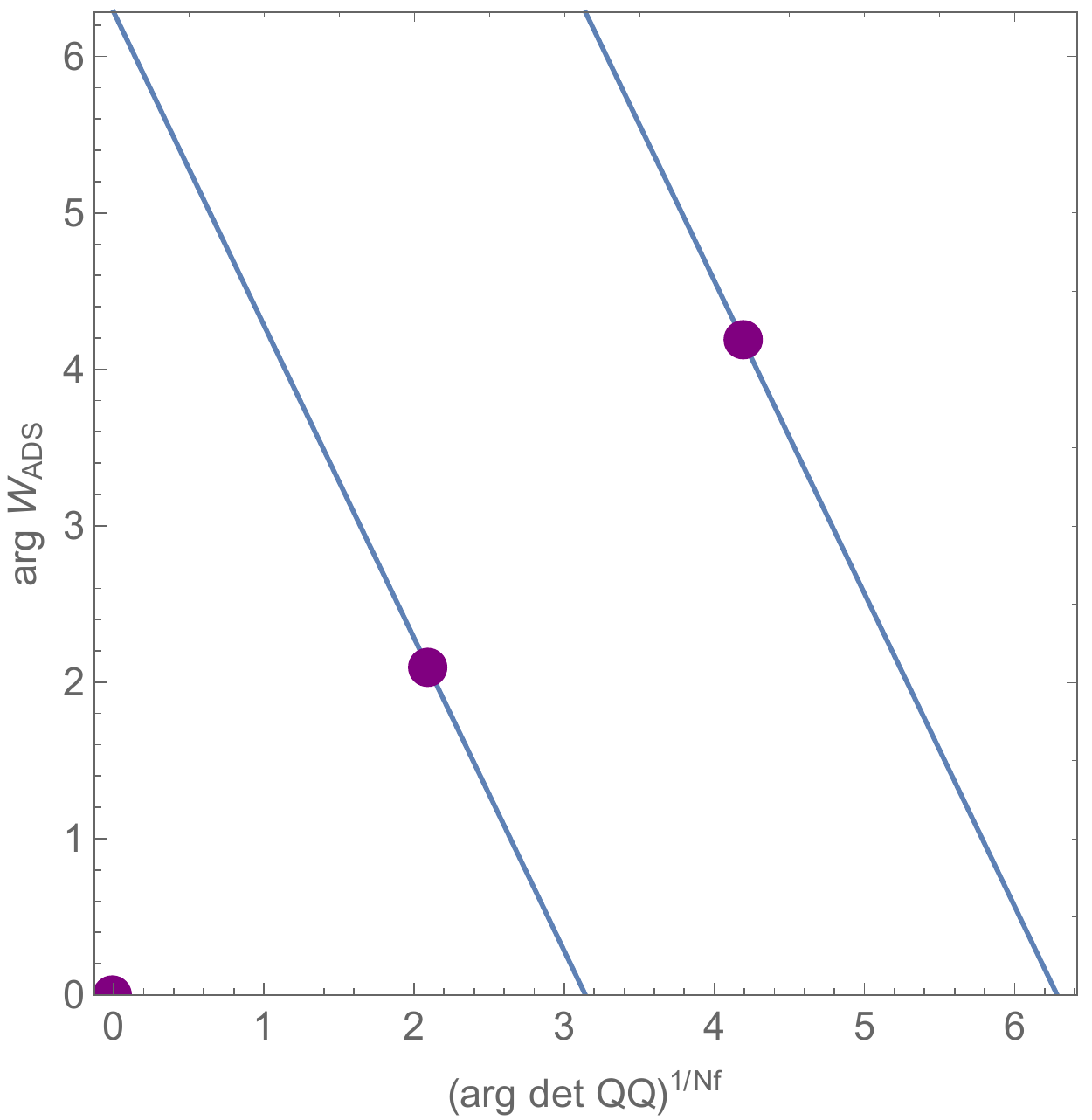}~~~~~
\includegraphics[width=0.5\linewidth]{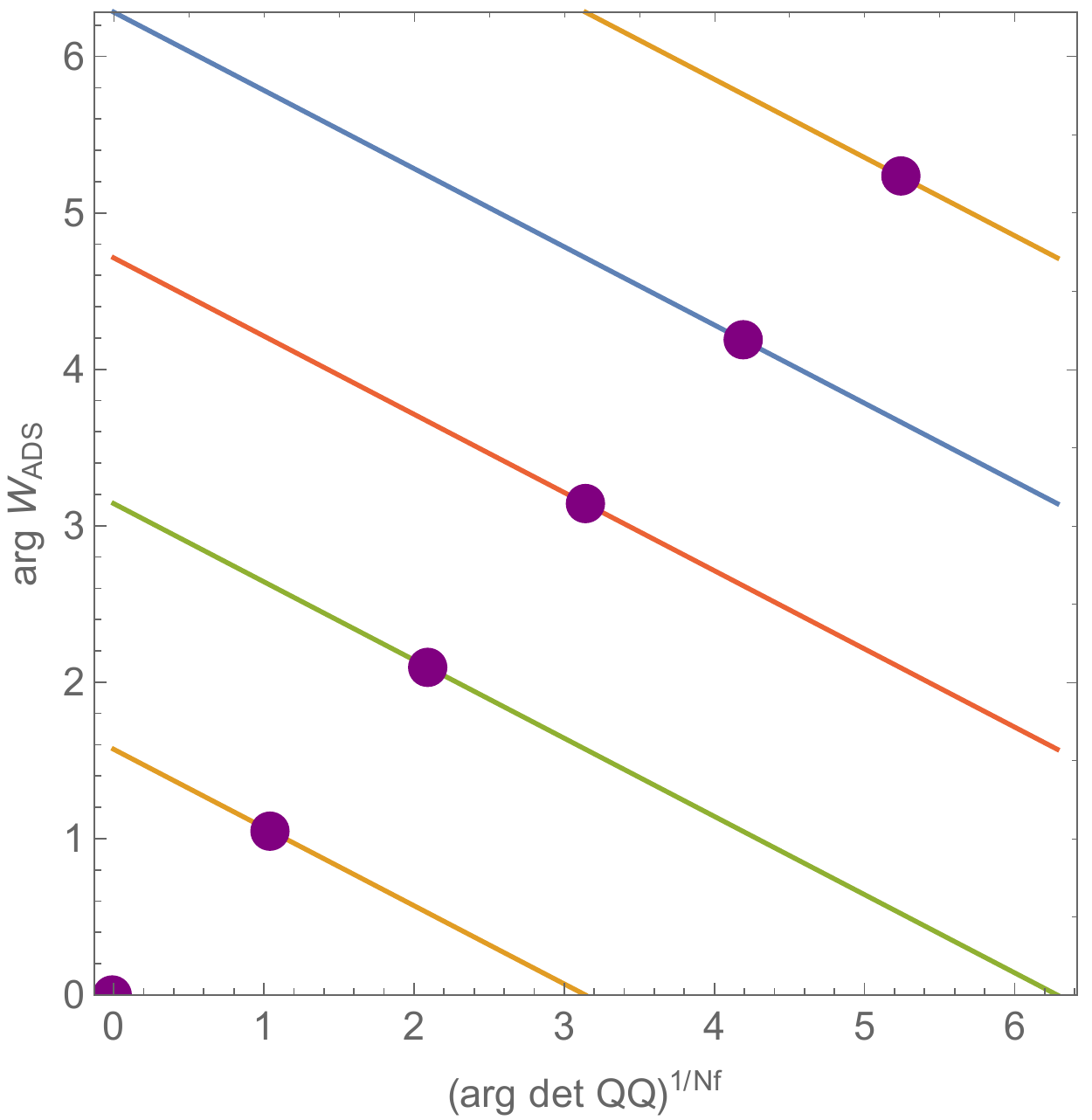}
\caption{Examples of branches and vacua in SQCD with $N_f<N$ flavors. Left: $N=3,N_f=2,\theta=0$. Right: $N=6,N_f=2,\theta=0$. Lines denote the subspace of the low-energy meson theory parametrized by $\eta^\prime\equiv(\arg\det Q\bar Q)^{1/N_f}$. For fixed $\eta^\prime$, there are $N-N_f$ branches of the gaugino bilinear ($\arg W_{ADS}$.) Wrapping  around $\Delta\eta^\prime=2\pi$ smoothly connects branch $k$ to branch $k-N_f \mod (N-N_f)$, leading to $\gcd(N_f,N-N_f)$ sets of branches that are disconnected in the $\eta^\prime$ direction. $N$ vacua are denoted by purple dots and are located along the line $\arg W=\eta^\prime$. The distance between neighboring vacua in the $\etap$ direction is $2\pi(N-N_f)/N$.   } 
\label{fig:thetazerovac}
\end{center}
\end{figure}

It is useful to visualize a subspace of the target space of the meson theory by restricting 
\begin{align}
(Q\bar Q)_{f\bar f}\rightarrow \delta_{f\bar f} v^2 e^{i\eta^\prime}\;.
\label{eq:etaprestriction}
\end{align}
The phase of superpotential (or the phase of the gaugino bilinear) is determined by the $\etap$,
\begin{align}
\arg W_{ADS} = \frac{-N_f\eta^\prime+2\pi k+\theta}{N-N_f}\;.
\label{eq:argllargqq}
\end{align}
From Eq.~(\ref{eq:argllargqq}), we can see that the domain of the $\etap$ is $g$ copies of $(p,q)$ torus knots wrapping around $\arg \lambda\lambda$, where
\begin{align}
g&=\gcd(N_f,N-N_f)\nonumber\\
p&=(N-N_f)/g\nonumber\\
q&=N_f/g\;.
\end{align}
Two examples are depicted in Fig.~\ref{fig:thetazerovac}.  For fixed $\eta^\prime$, there are $N-N_f$ branches of the gaugino bilinear, while for fixed $\arg \lambda\lambda$, there are $N_f$ branches for the $\eta^\prime$. 

Motion in ``pion-like" directions, where $Q\bar Q$ is not proportional to the identity, is also important.  In a convenient basis, we have
\begin{align}
e^{2\pi i /N_f}&=e^{2\pi i T_{-1}/N_f}\nonumber\\
T_{-1}= {\rm diag}(&1,\dots,1,1-N_f)\;.
\label{eq:piondirec}
\end{align}
Thus motion by $2\pi/N_f$ in the $\pi_{-1}$ direction of the pseudogoldstone manifold connects different $\eta^\prime$ branches at fixed $\arg\lambda\lambda$. We will use the $\pi_{-1}$ and $\etap$ directions to make ansatzes for domain wall trajectories. 

\subsection{Vacua}
The meson vacua lie at $N$ values of $\etap$, with
\begin{align}
\frac{N\langle\eta^\prime\rangle -\theta-2\pi k}{N-N_f} = 2\pi\omega\;,\;\;\;\omega\in \mathbf Z
\end{align}
and
\begin{align}
v=\Lambda^{\frac{3N-N_f}{2N}}m^{\frac{N_f-N}{2N}}\;.
\label{eq:vvac}
\end{align} 
Requiring $\eta^\prime\in[0,2\pi)$, we have
\begin{align}
\langle\eta^\prime\rangle&=\frac{2\pi\omega (N-N_f)+2\pi k+\theta}{N} \\
&=\frac{2\pi n+\theta}{N}\;,\;\;\;\;\;n=0\dots N-1\;,
\label{eq:nvac}
\end{align}
and the phase of the superpotential/gaugino bilinear is
\begin{align}
\arg\langle W\rangle=\langle\eta^\prime\rangle\;.
\label{eq:Wetavac}
\end{align}
The distance between two neighboring vacua in the $\eta^\prime$ direction is 
\begin{align}
\Delta\langle\eta^\prime\rangle=\frac{2\pi (N-N_f)}{N}
\end{align}
In the two examples shown in Fig.~\ref{fig:thetazerovac}, the vacua are denoted by purple dots.

\section{$CP$ in Softly Broken SQCD}
\label{sec:CP}
In the supersymmetric theory, we can always use the anomalous symmetries to rotate away $\theta$ and $\arg \det m$. Then there always exists a definition of $CP$ which is not spontaneously broken.

In the presence of soft breaking terms, there are physical vacuum angles, and $CP$ can be spontaneously broken when some combinations of them are equal to $\pi$. We add to the theory small soft masses of the form\footnote{We could add $m_Q^2|Q|^2+m_{\bar Q}^2|\bar Q|^2$ soft masses as well, but will not do so here for simplicity; in Sec.~\ref{sec:qcdadj} we will comment on decoupling the scalars with large soft masses.}
\begin{align}
V_{soft}=m_\lambda \langle\lambda\lambda\rangle + m B\,\TR(Q\bar Q)+c.c.\;.
\label{eq:Vsoft}
\end{align}
We will assume that $m_\lambda, B\ll m$. In this limit the vacua are still approximately by Eq.~(\ref{eq:vvac}),(\ref{eq:nvac}), and once we have defined the $CP$ symmetry, we can identify which vacua are related by this definition of $CP$ in the SUSY limit.

When the soft masses are nonzero, the theory contains two invariant phases, which we can take to be
\begin{align}
\bar\theta_A&=\theta+N\arg m_\lambda + N_f \arg m\nonumber\\
\bar\theta_B&=\theta+N\arg B + N_f \arg m\;.
\end{align}

It is convenient to take $m$ and $m_\lambda$ to be real and positive, which can be achieved in general using the anomalous $U(1)_A$ and $U(1)_R$ symmetries. Then the remaining phases are $\theta$ and $\arg B$.\footnote{Invariant phases can be restored in what follows with the substitutions $\theta\rightarrow\bar\theta_A,~\arg B\rightarrow (\bar\theta_B-\bar\theta_A)/N$.} 

For general $\theta,~\arg B$, there is one unique global vacuum and there are no static domain walls. When these phases are 0 or $\pi$, however, a $CP$ symmetry can be defined, and in some cases it is spontaneously broken, leading to static walls. For either value of $\arg B$, $CP$ acts on the ADS effective theory as 
\begin{align}
\theta&=0: ~~~~(Q\bar Q,~k)\rightarrow(Q\bar Q^\dagger,~-k)\nonumber\\
\theta&=\pi: ~~~~(Q\bar Q,~k)\rightarrow(Q\bar Q^\dagger,~-k-1)\;.
\end{align}
Correspondingly, in the supersymmetric limit, this $CP$ symmetry acts on the $n$-vacua of Eq.~(\ref{eq:nvac}) as
\begin{align}
\theta&=0: ~~~~n\rightarrow -n\nonumber\\
\theta&=\pi: ~~~~n\rightarrow -n-1\;.
\label{eq:nCP}
\end{align}

The realization of $CP$ is determined by which vacua remian global minima in the presence of soft breaking. 
At leading order in the soft parameters, the changes in the vacuum energies are determined entirely by $\langle n| V_{soft}|n\rangle$, where $n$ are the $N$ vacua in the SUSY limit.  Using Eqs.~(\ref{eq:ll}) and~(\ref{eq:Wetavac}) and assuming $\arg B$ is $0$ or $\pi$, we have
\begin{align}
\langle V_{soft}\rangle
=Y\cos\left(\frac{2\pi n+\theta}{N}\right)
\end{align}
where
\begin{align}
 Y&\equiv
-m_\lambda |\langle\lambda\lambda\rangle|+ m B \, |\TR(\langle Q\bar Q\rangle)| \;\nonumber\\
&=\left(-32\pi^2 m_\lambda+ N_fB\right)\Lambda^{\frac{3N-N_f}{N}}m^{\frac{N_f}{N}}\;.
\label{eq:vsoftvac}
\end{align}
Since we have used the anomalous symmetries to make $m$ and $m_\lambda$ real and positive, $Y$ is real. The realization of $CP$ is as follows:
\begin{itemize}
\item If $Y<0$,  then $CP$ is spontaneously broken iff $\theta=\pi$. The vacua are  $n=0,\,N-1$. 
\item If $Y>0$, $CP$ is spontaneously broken either if $\theta=0$ and $N$ is odd, in which case the vacua are $n=(N\pm 1)/2$, or if $\theta=\pi$ and $N$ is even, in which case the vacua are $n=(N-1\pm1)/2$.  
\end{itemize}
We can summarize these various cases by saying that $CP$ is spontaneously broken iff
\begin{align}
\alpha=\pi\;,
\label{eq:alphapi}
\end{align}
where 
\begin{align}
\alpha\equiv \arg\left[\left(-Y\right)^N e^{i\theta}\right]\;.
\label{eq:alpha}
\end{align}
Note that due to the presence of two invariant phases in the theory, $\bar\theta_A=\pi$ is neither necessary nor sufficient to spontaneously break $CP$. 

We can also check two limits, $B=0$ and $m_\lambda=0$. In each case there is only one angle, $\bar\theta_A$ and $\bar\theta_B$ respectively. In these limits the condition $\alpha=\pi$ reduces to  $\bar\theta_A=\pi$ or $\bar\theta_B+N\pi=\pi$ when written in terms of the invariant angles.

\section{$CP$ Domain Walls}
\label{sec:walls}
When $CP$ is spontaneously broken, we expect to find domain wall solutions connecting the vacua. For small soft breaking parameters, some walls may be approximately those of the supersymmetric theory, in which case we can study them in that context. Other walls may depend critically on the presence of the soft breakings. In addition, in the former case the walls may be approximately BPS. The BPS equations and a few of the wall properties are reviewed in the appendix. 

In the simplest case, walls connect nearest neighbors in the $\etap$ direction, preserving the $SU(N_f)_V$ global symmetry. However, following Eq.~(\ref{eq:piondirec}), walls can also traverse a combination of pion-like and $\etap$-like directions, homotopically inequivalent to the pure-$\etap$ wall in the IR theory.  When $g>1$, some vacua can only be connected in directions that involve a pion, because some gaugino branches are not connected in the $\etap$ direction (see, for example, the right-hand panel of Fig.~(\ref{fig:thetazerovac}).) To accommodate both possibilities, we can take $Q\bar Q$ to have the form
\begin{align}
Q\bar Q = {\rm diag}\left(\rho_1^2 e^{i\phi_1},\dots,\rho_1^2 e^{i\phi_1},\rho_2^2 e^{i\phi_2}\right)\;.
\end{align}
This is only an ansatz, and in general other pion-like trajectories may exist, but it will be sufficient for our purposes. The $\phi_{1,2}$ phase basis is related to $\etap,\pi_{-1}$ ($\pi_{-1}$ is associated with the generator given in Eq.~(\ref{eq:piondirec})) as
\begin{align}
\etap=\frac{(-1+N_f)\phi_1+\phi_2}{N_f}\;,\;\;\;\;
\pi_{-1}=\frac{\phi_1-\phi_2}{N_f}\;.
\end{align}

Below we consider three representative cases for $N$ and $N_f$ in greater detail. For simplicity, we focus on the spontaneous $CP$ breaking scenario $\arg B=\pi$, where, as discussed in the previous section, $CP$ is always broken at $\theta=\pi$ and the minima are at $n=0,N-1$ in the SUSY limit.

\subsection{$\gcd(N,N_f)>1$}
$\gcd(N,N_f)>1$ implies $g=\gcd(N,N-N_f)>1$, in which case, as we have discussed, adjacent vacua cannot be connected by motion purely in the $\etap$ direction in field space. Therefore, there are no trivial $\etap$ walls connecting the $CP$ vacua in these theories. However, pion-like walls still exist. 
\begin{figure}[t!]
\begin{center}
\includegraphics[width=0.5\linewidth]{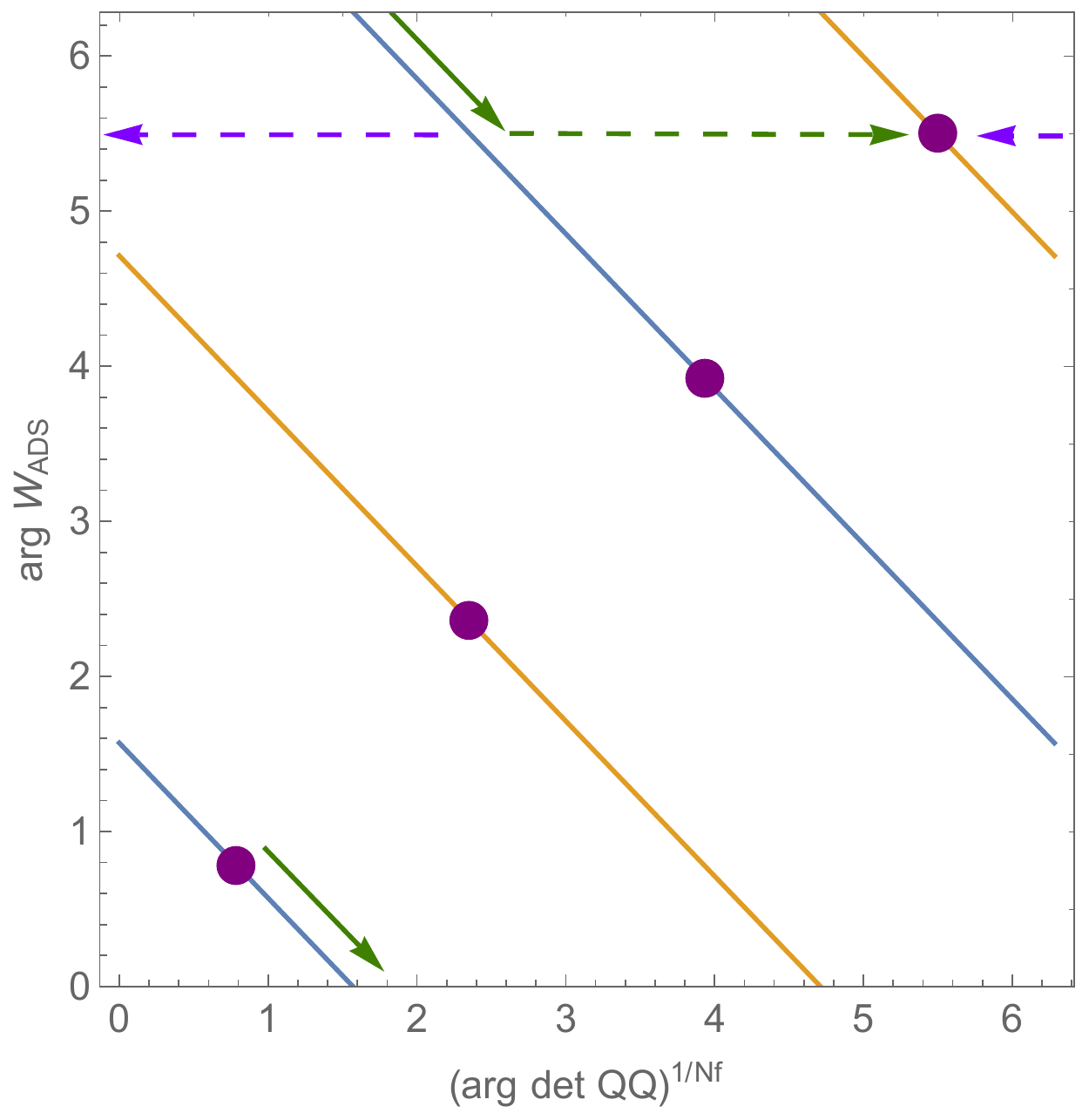}
\caption{Cartoon of domain wall trajectories between nearest neighbor vacua in $N_f=2,N=4$. Arrows denote different possible wall trajectories in mixed $\etap$-pion directions (solid corresponds to the $\etap$ direction; dashed corresponds to motion ``out of the page" in the $\pi_{-1}$ direction.) The $CP$ conjugate vacua cannot be connected by motion purely in the $\etap$ direction. } 
\label{fig:Nf2N4cartoon}
\end{center}
\end{figure}

\begin{figure}[t!]
\begin{center}
\includegraphics[width=0.45\linewidth]{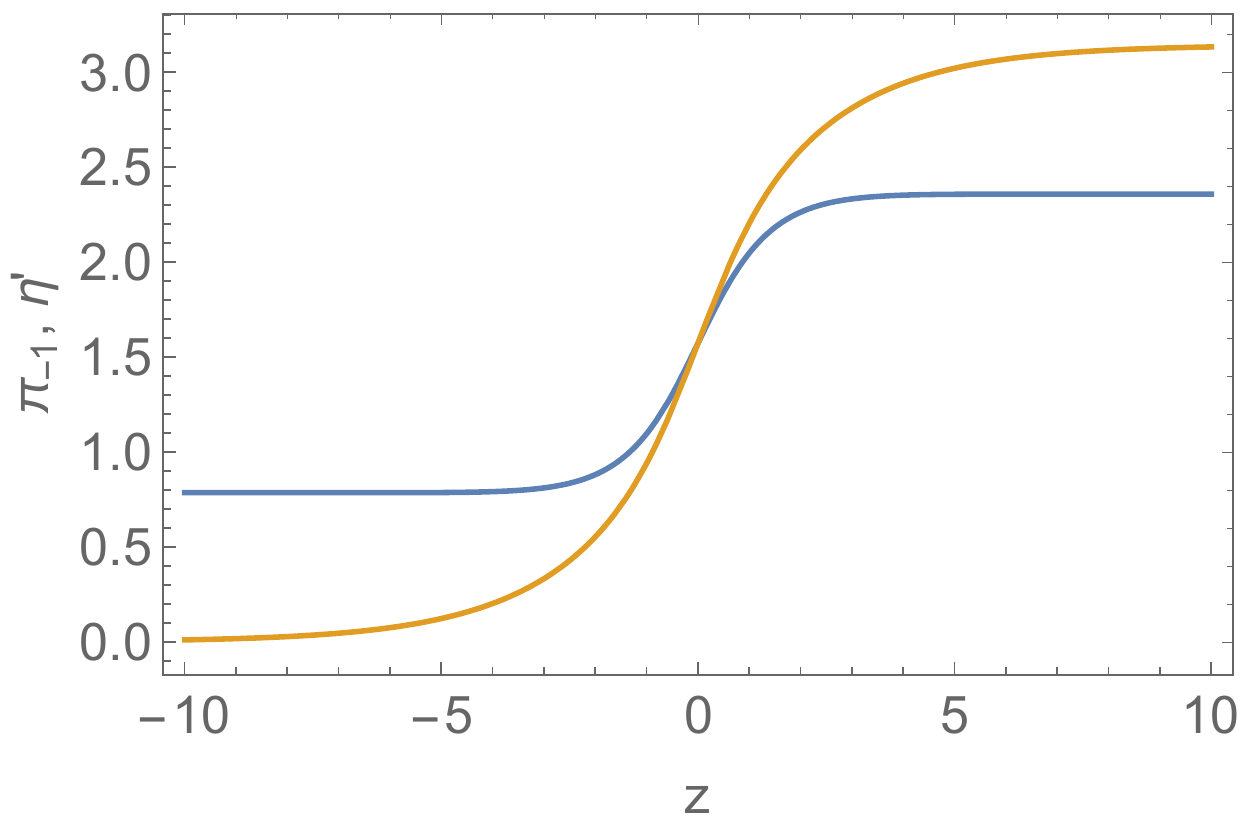}
\includegraphics[width=0.45\linewidth]{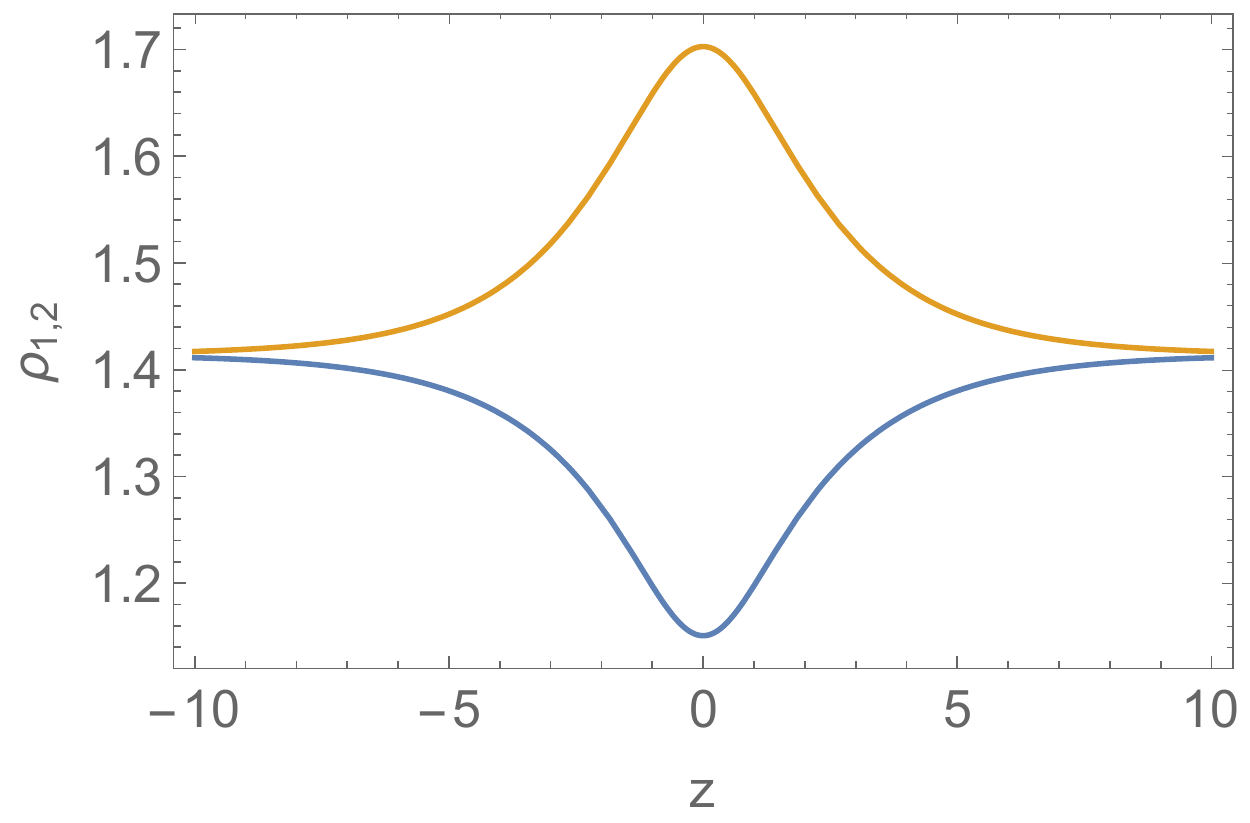}
\caption{Phase and modulus profiles of an $SU(N_f)_V$-breaking BPS domain wall in $N=4, N_f=2$, corresponding to the green path in Fig.~\ref{fig:Nf2N4cartoon}. In this case, since $\gcd(N,N_f)>1$, only $SU(N_f)_V$-breaking walls are present between the $CP$ vacua. For definiteness we have set $\Lambda/m=4$. In the left-hand panel, the orange (blue) curve corresponds to the $\pi_{-1}$ ($\etap$). In the right-hand panel, the orange (blue) curve corresponds to $\rho_1$ ($\rho_2$).}
\label{fig:Nf2N4pion}
\end{center}
\end{figure} 

A cartoon of pion-like trajectories between $CP$ conjugate vacua is shown in Fig.~\ref{fig:Nf2N4cartoon} for $N_f=2,N=4$. We find a BPS solution numerically in the SUSY limit, shown in Fig.~\ref{fig:Nf2N4pion}. For small soft breakings this is the leading-order approximation to the minimal tension wall connecting the $CP$ conjugate vacua. When we discuss the case $N_f=N-1$ below, we will show that the leading-order change in the wall tension from soft terms is determined entirely by the BPS trajectory between the supersymmetric vacua.

The absence of $\etap$ walls for $g>1$ also reflects the anomaly structure of the microscropic theory, discussed further in Sec~\ref{sec:anomalies} below. Nonsupersymmetric theories with $g>1$ have an 't Hooft anomaly at $\theta=\pi$ between $CP$ and a discrete subgroup of the flavor symmetry intertwined with the center of the gauge group~\cite{qcdcpdw}. At low energies, the latter is unbroken, so the anomaly is saturated by the spontaneous breaking of $CP$. The same symmetries and anomaly are also present in the softly-broken supersymmetric theory, and $CP$ is spontaneously broken at low energies. A discrete version of anomaly inflow requires domain walls to possess nontrivial infrared degrees of freedom. The $\etap$ walls have only a translation zero mode at low energies and cannot saturate the anomaly, so they are forbidden.

\subsection{$N_f<N-1$, $\gcd(N,N_f)=1$}

\begin{figure}[t!]
\begin{center}
\includegraphics[width=0.5\linewidth]{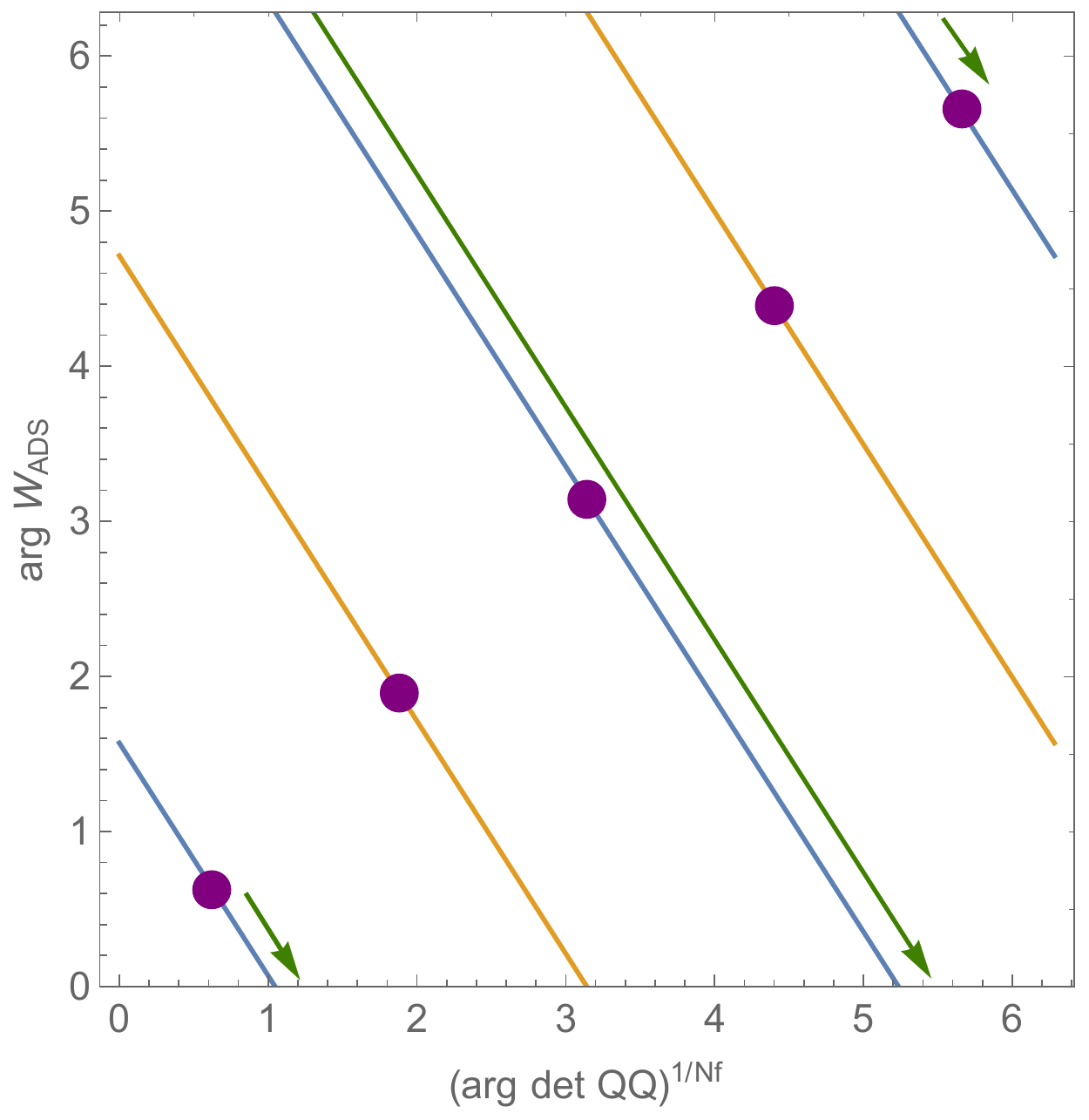}
\caption{
Cartoon of the domain wall trajectory between $CP$ vacua for $N_f=3,N=5$. Arrows denote a wall trajectory in the $\etap$ direction, passing through an intermediate vacuum at $\etap=\pi$. (Mixed $\pi_{-1}-\etap$ trajectories not shown.) 
} 
\label{fig:Nf3N5cartoon}
\end{center}
\vskip 2.2cm
\end{figure} 
\begin{figure}[h!]
\begin{center}
\includegraphics[width=0.46\linewidth]{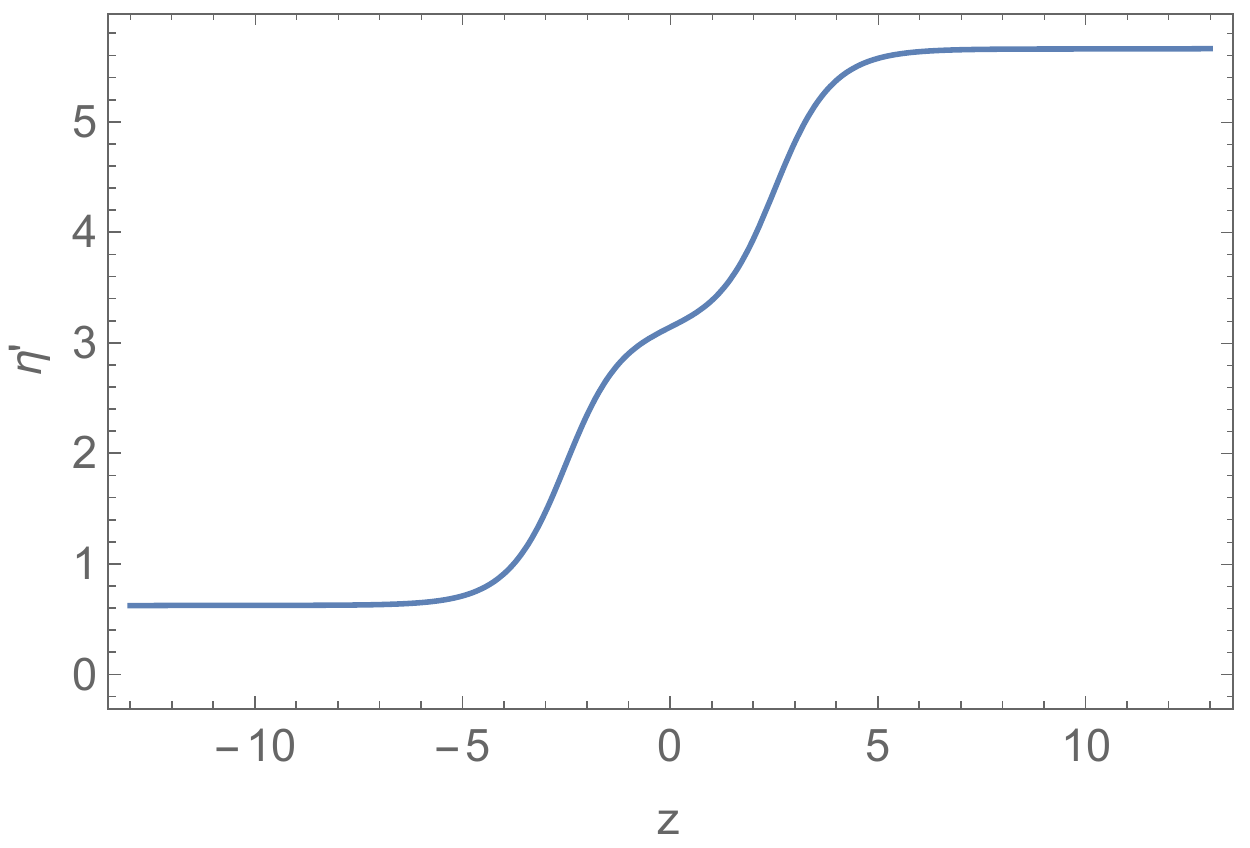}
\includegraphics[width=0.46\linewidth]{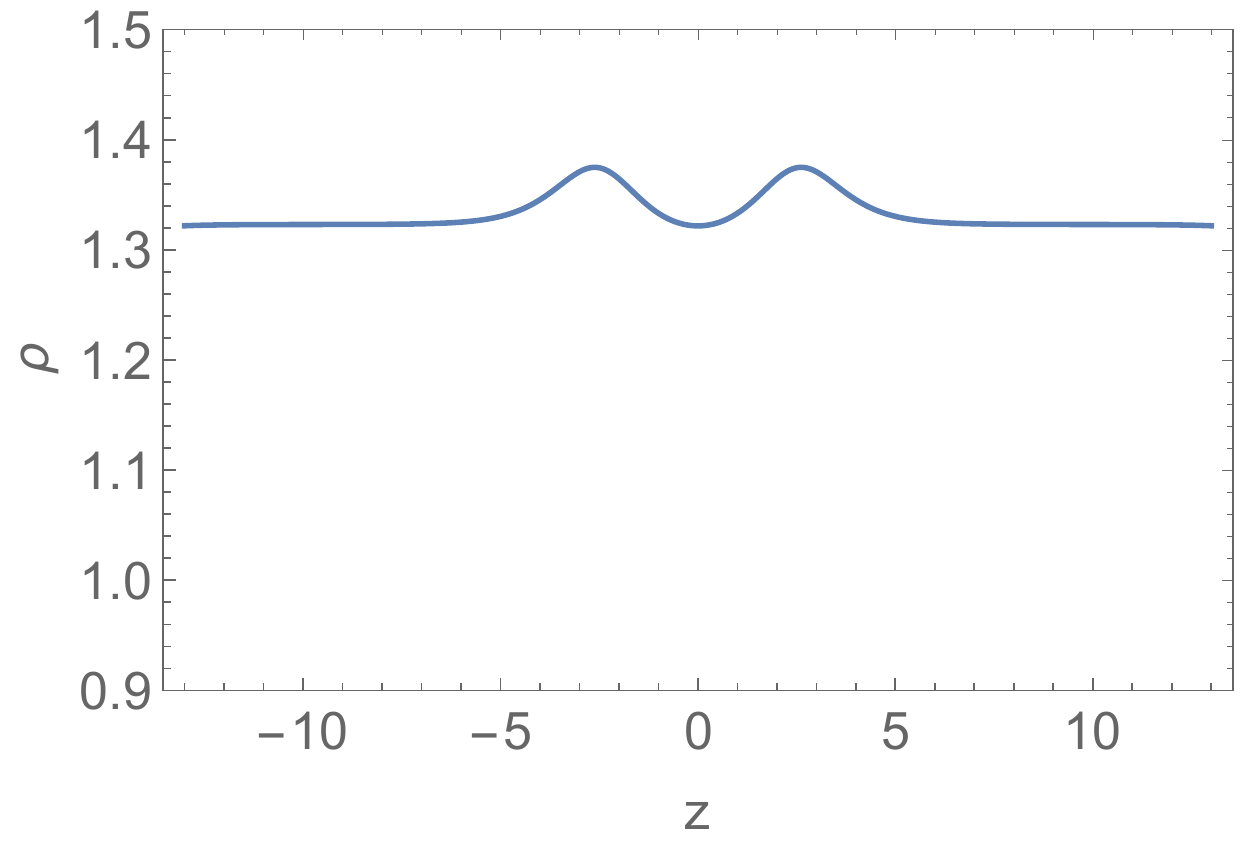}
\caption{Phase and modulus profile of a domain wall in $N=5, N_f=3$, corresponding to the green trajectory in Fig.~\ref{fig:Nf3N5cartoon}. In this case, since $\gcd(N_f,N-N_f)=1$, an $SU(N_f)_V$-preserving wall is present between the $CP$ vacua. However, it is not BPS, and exists only for finite SUSY-breaking. We have taken $\Lambda/m=4$ and $B=-1/50$.} 
\label{fig:Nf3N5etap}
\end{center}
\end{figure} 

When $g=1$  and $N_f\neq N-1$ (and $N_f\neq 1$, a trivial case we do not consider explicitly), the $CP$ vacua lie along branches that are smoothly connected in the $\etap$ direction. However, such a trajectory must first pass through one or more intervening vacua. The presence of these vacua prevents, in the SUSY limit, the existence of BPS walls. While we could contemplate superposing widely separated BPS walls connecting the vacua as they are encountered in the $\etap$ direction, such walls would neither be exactly static nor BPS.

The problem is that the intervening vacua are degenerate with the $CP$ vacua in the SUSY limit. With small SUSY breaking, however, the former are lifted. Now a domain wall can pass through the lifted vacua in finite ``time" and make its way to the $CP$ conjugate minimum.

This expectation is confirmed numerically in the example $N_f=3,N=5$. The trajectory is sketched in Fig.~\ref{fig:Nf3N5cartoon}, showing that an $\etap$ wall connecting $CP$ vacua must pass through one additional metastable vacuum on the way.  No BPS or non-BPS solutions were found in the SUSY limit. Adding a small soft mass (a $B$-term for simplicity, with $\arg B=\pi$), we find a solution, shown in Fig.~\ref{fig:Nf3N5etap}. As expected, the trajectory looks qualitatively like two superposed BPS walls connecting the $CP$ vacua to the vacuum at $\etap=\pi$. For arbitrarily small soft breaking, the wall spends more and more ``time" near the latter vacuum, pushing the superposed walls farther apart toward infinity.

\subsection{$N_f=N-1$}

\begin{figure}[t!]
\begin{center}
\includegraphics[width=0.5\linewidth]{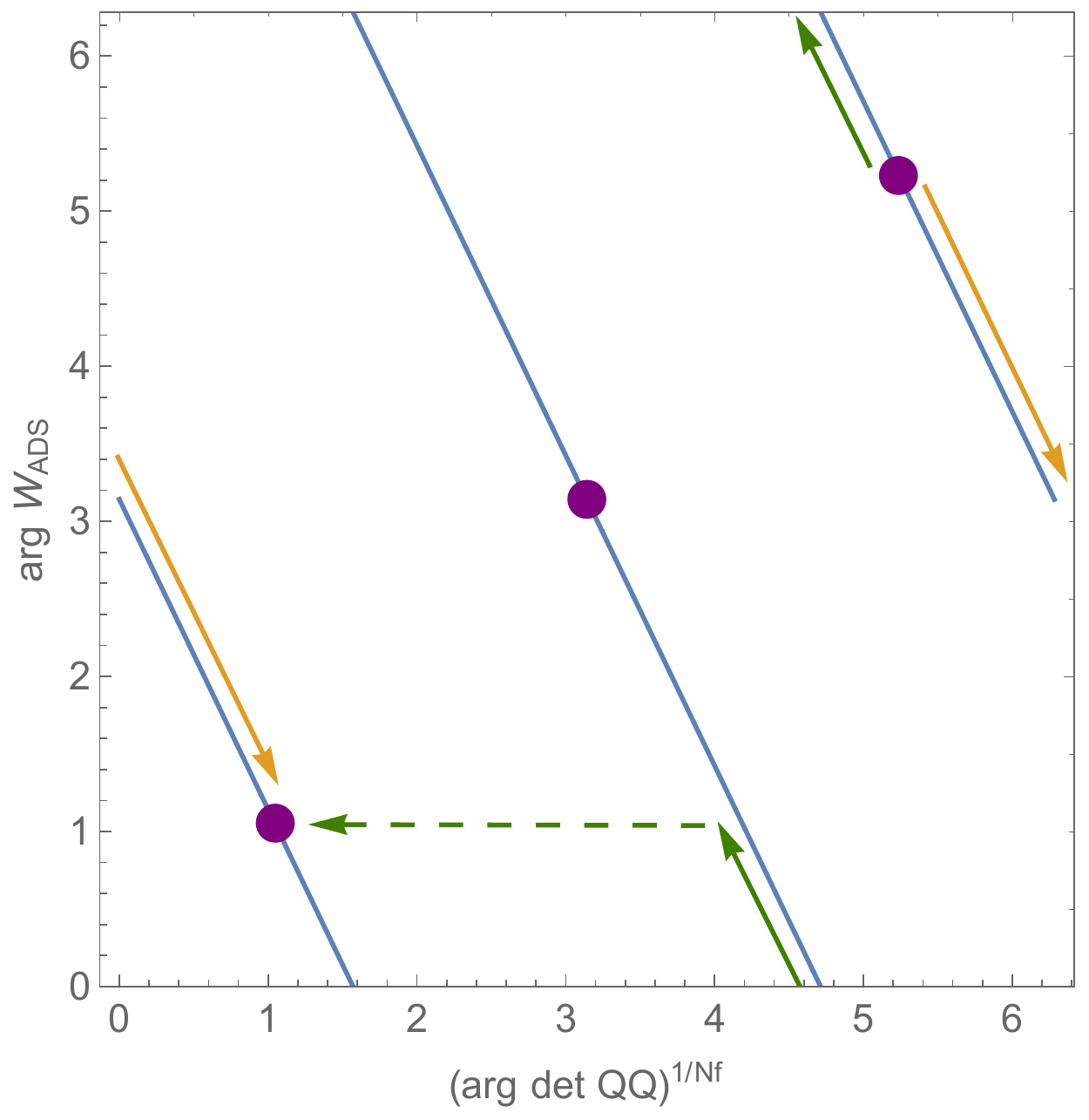}
\caption{Cartoon of domain wall trajectories between nearest neighbor vacua in $N_f=N-1$. Orange arrows denote a purely $\etap$-type trajectory. Green arrows denote a possible mixed $\etap$-pion direction (solid corresponds to the $\etap$ direction; dashed corresponds to motion ``out of the page" in the $\pi_{-1}$ direction, which connects $\etap$ branches according to Eq.~(\ref{eq:piondirec}).)} 
\label{fig:Nf2N3cartoon}
\end{center}
\end{figure}

\begin{figure}[t!]
\begin{center}
\includegraphics[width=0.46\linewidth]{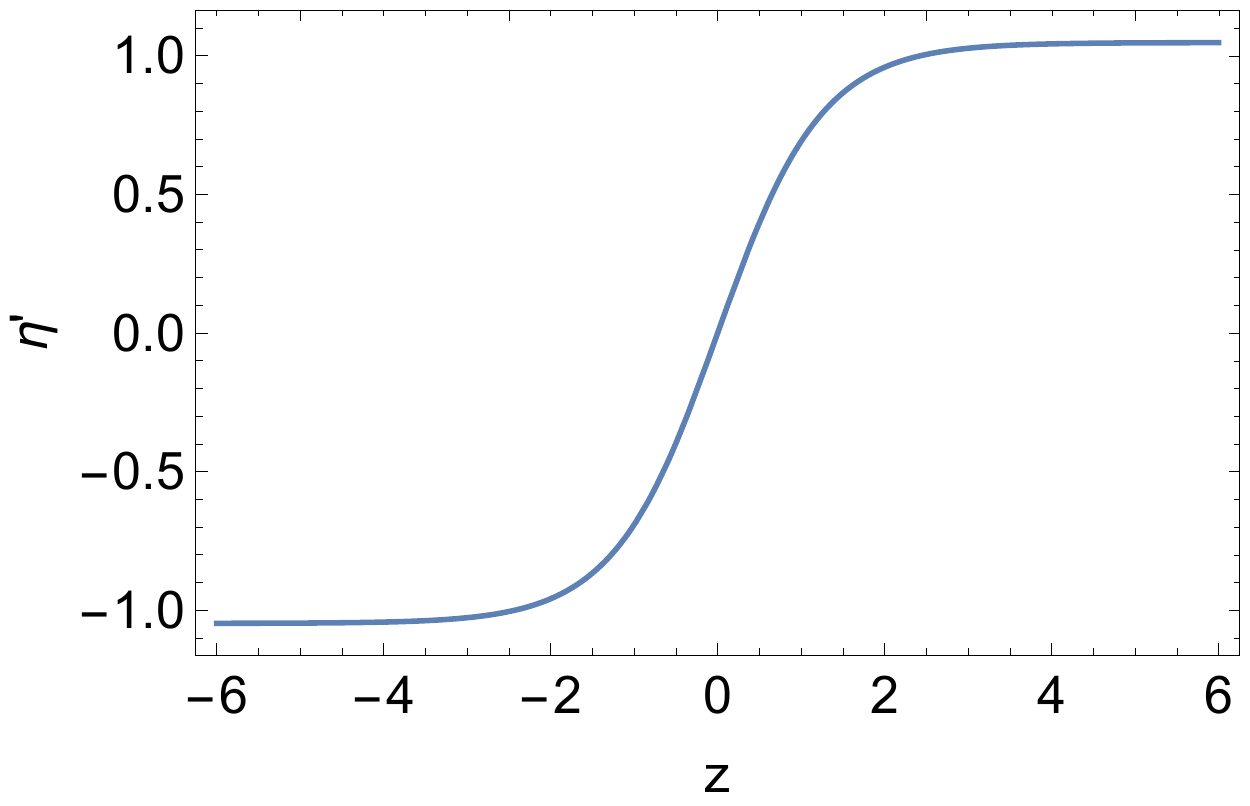}
\includegraphics[width=0.46\linewidth]{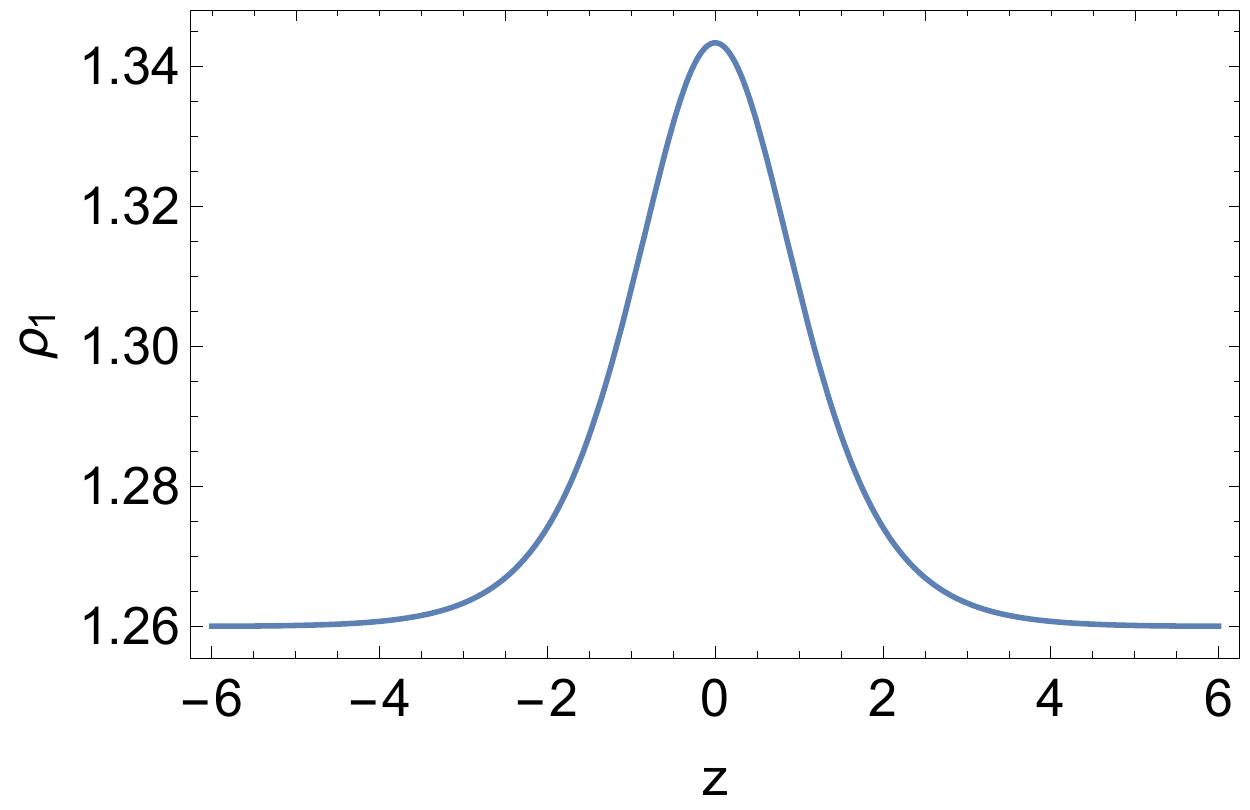}
\caption{Phase and modulus profile of an $SU(N_f)_V$-preserving BPS domain wall in $N=3, N_f=2$, corresponding to the orange trajectory in Fig.~\ref{fig:Nf2N3cartoon}. Parameters are as in Fig.~\ref{fig:Nf2N4pion}.} 
\label{fig:Nf2N3etap}
\end{center}
\end{figure} 
\begin{figure}[h!]
\begin{center}
\includegraphics[width=0.46\linewidth]{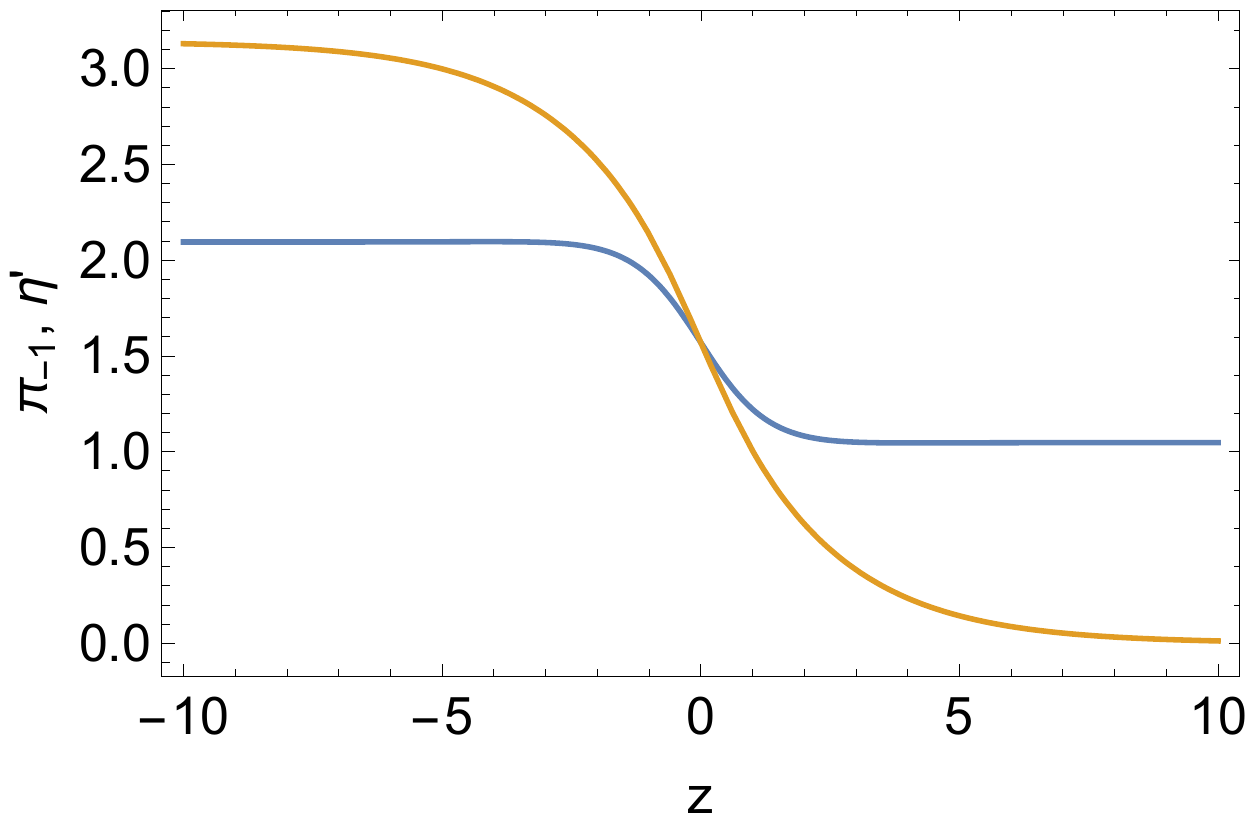}
\includegraphics[width=0.46\linewidth]{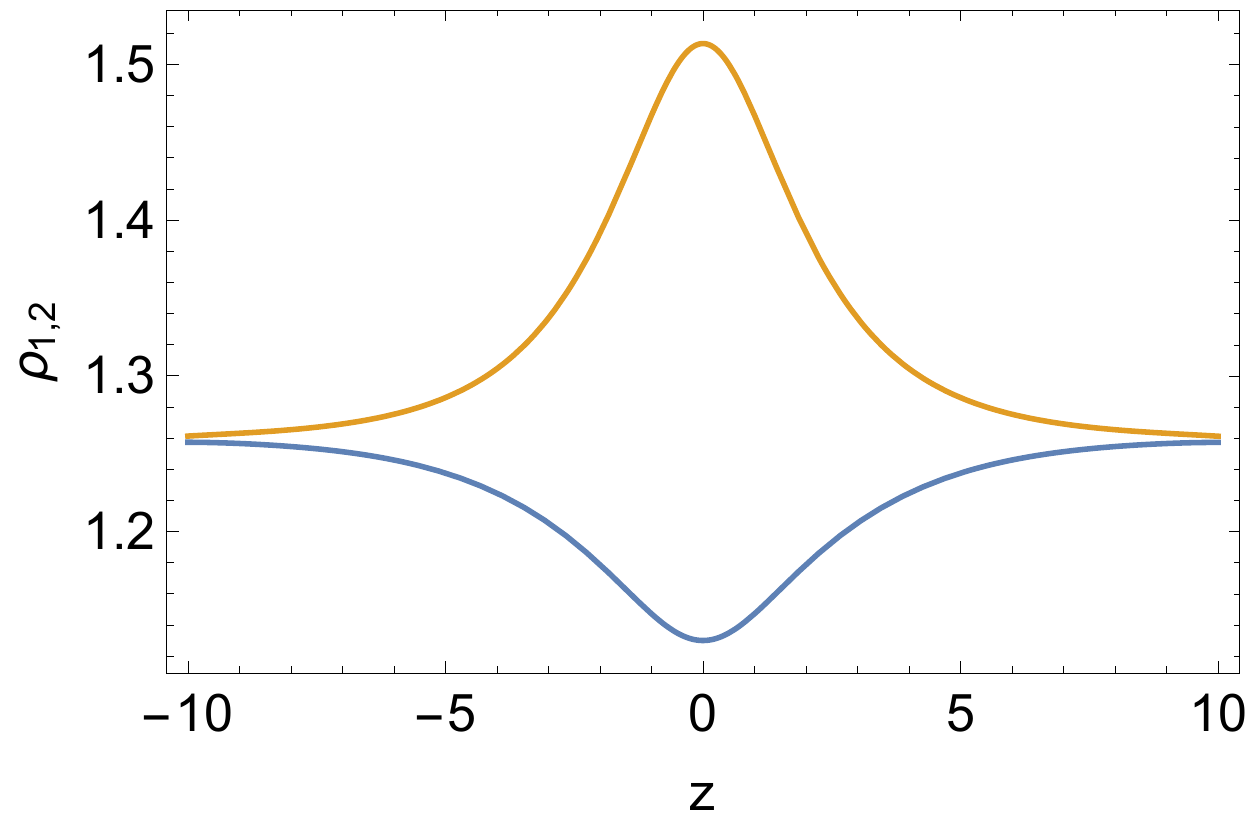}
\caption{Phase and modulus profile of an $SU(N_f)_V$-breaking BPS domain wall in $N=3, N_f=2$, corresponding to the green trajectory in Fig.~\ref{fig:Nf2N3cartoon}. Parameters and curves are as in Fig.~\ref{fig:Nf2N4pion}.} 
\label{fig:Nf2N3pion}
\end{center}
\end{figure} 

For $N_f=N-1$, the $CP$ vacua are at $n=0,N-1$ in the SUSY limit, and are nearest neighbors in the $\etap$ direction. Domain walls between these vacua can run either purely in the $\etap$ direction or in a mixed $\etap-\pi_{-1}$ direction. In the SUSY limit, if the walls are BPS, they are of equal tension.

 A cartoon for $N=3$, $N_f=2$ is shown in Fig.~\ref{fig:Nf2N3cartoon}. BPS walls for this case were studied in~\cite{smilga}. We construct solutions numerically from the BPS equations and show the trajectories for a pure $\etap$ wall and a mixed $\etap$-$\pi_{-1}$ wall  in Figs.~\ref{fig:Nf2N3etap} and~\ref{fig:Nf2N3pion} respectively. The former preserves the vectorlike flavor symmetry, while the latter partially spontaneously breaks it.

These BPS domain walls are the leading approximation to the domain walls connecting the $CP$ conjugate vacua in the presence of small SUSY breaking. For small soft parameters, the changes in the tensions can be computed from the BPS walls, without knowledge of the modifications to the trajectories from SUSY breaking.  At first order in the soft masses, the only contribution to the change in the wall tensions is the integral over the soft terms evaluated on the supersymmetric walls,
\begin{align}
\Delta T = \int\,dz\, V_{soft}(Q\bar Q_{BPS}) \;.
\label{eq:DeltaT}
\end{align}
All other changes to the trajectory and tension are of higher order. Therefore, the numerical solutions in Figs.~\ref{fig:Nf2N3etap} and~\ref{fig:Nf2N3pion} are sufficient to work out the leading changes to the tensions. In general one wall becomes metastable, having higher tension than the other, and we can exhibit a phase diagram.

\begin{figure}[t!]
\begin{center}
\includegraphics[width=0.5\linewidth]{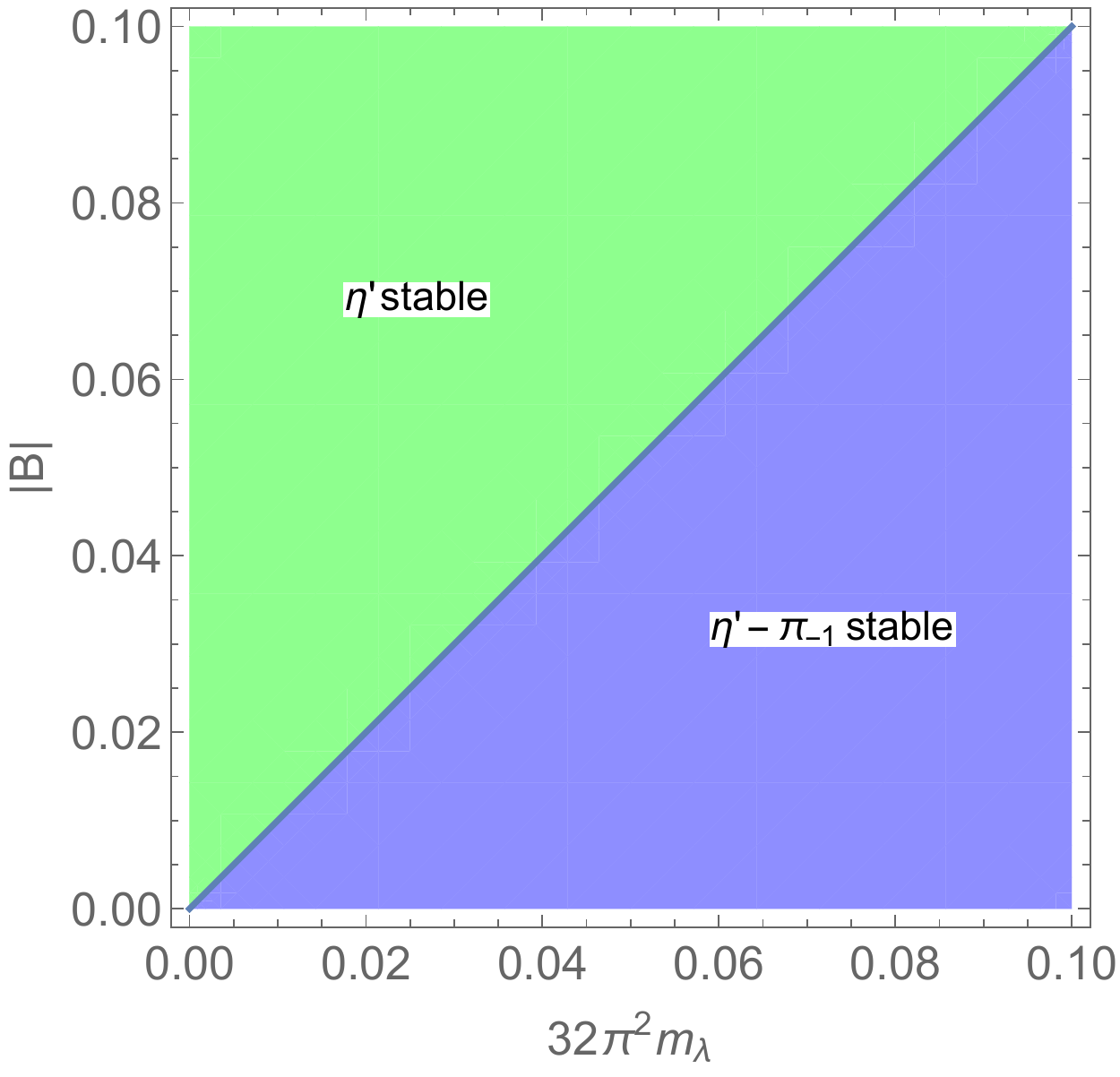}
\caption{Regions on the soft mass parameter space where different $CP$-wall trajectories are stable for $\theta=\arg B=\pi$. In the upper left, the stable wall is $SU(N_f)_V$ preserving. In the lower right, the stable wall partially breaks the flavor symmetry.} 
\label{fig:Nf2N3phasediag}
\end{center}
\end{figure} 

To gain some insight into the wall phase diagram, note that the BPS walls admit a constant of motion
\begin{align}
E={\rm Im}(e^{i\delta}W)
\end{align}
where $\delta$ is found by computing $E$ in the two vacua,
\begin{align}
\delta=\frac{\pi}{2}-\left(\frac{\langle\etap\rangle_1+\langle\etap\rangle_2}{2}\right)
\end{align}
For the $CP$ vacua $n=0,N-1$, $\delta\rightarrow\pi/2$. Therefore $E={\rm Re}(W)$ for all walls connecting these vacua.

Furthermore, when $B=-32\pi^2 m_\lambda$, $V_{soft}\propto{\rm Re}(W)$. Therefore, for these parameters, tensions along all different wall trajectories remain equivalent at leading order in the soft masses. In fact, this line demarcates a phase transition across which the $CP$ wall changes its field space configuration. 

We integrate the soft potential numerically over the BPS solutions for the sample case $N=3, N_f=2$. Fig~\ref{fig:Nf2N3phasediag} shows the resulting phase diagram for the $CP$ domain walls. The low-energy theory on the walls undergoes a first-order phase transition across the line $V_{soft}\propto{\rm Re}(W)$.

\subsection{'t Hooft Anomaly}
\label{sec:anomalies}
As mentioned in the introduction, the $CP$ anomaly found in~\cite{qcdcpdw} in nonsupersymmetric QCD also applies to SQCD with small soft masses. We briefly review the derivation of this anomaly. We turn on background gauge fields for the $U(N_f)_V$ flavor symmetries and possible counterterms for their topological charges. The gauge fields are then those of $SU(N)\times SU(N_f)\times U(1)/(\mathbb{Z}_N\times\mathbb{Z}_{N_f})$.  The quotient by $\mathbb{Z}_N$ arises because the action of this subgroup of the flavor symmetries on matter fields in the fundamental representation can be undone by a transformation in the center of the gauge group. 

On $T^4$, we can consider twisted gauge transformations. If the  transformation is twisted by an element of the center, it is not a gauge transformation, but it is a global symmetry of the action in the theory without fundamental flavors. In this theory we can gauge this discrete center symmetry, or introduce background gauge fields for it~\cite{Aharony:2013hda,Kapustin:2014gua,Gaiotto:2014kfa}. Either amounts to considering gauge field configurations carrying 't Hooft fluxes~\cite{tHooft:1979rtg}. In these configurations the gauge fields differ across one cycle by a gauge transformation which is itself twisted around another cycle by a center transformation. Combinations of electric and magnetic fluxes of this type can carry fractional topological charge of order $1/N$.

Boundary conditions for fields in the fundamental representation explicitly break the center symmetry, so it cannot be gauged by itself. However, combining center with twisted flavor transformations  preserves the boundary conditions. Therefore we can still consider 't Hooft fluxes as long as they are correlated between the original gauge symmetry and background gauge fields for the flavor symmetry.

At $\theta=\pi$, the ordinary topological charge plus counterterms is
\begin{align}
\frac{\pi}{16\pi^2}\int\TR(F\tilde F)+\frac{\theta_f}{16\pi^2}\int\TR(F_f\tilde F_f)+\frac{\theta_B}{16\pi^2}\int F_B\tilde F_B\;.
\end{align}
Each topological charge is integral for $SU(N)\times SU(N_f)\times U(1)$ gauge fields. Then the action is $CP$ invariant if $\theta_{f,B}$ are $0$ or $\pi$. However, with 't Hooft fluxes, it might be impossible to chose $\theta_{f,B}$ so that the action is $CP$ invariant. The original theory then possess an 't Hooft anomaly between $CP$ and the intertwined flavor-center global symmetry.

By considering different simple choices for the fluxes with ${\cal O}(1/N)$ and ${\cal O}(1/N_f)$ topological charges, it is straightforward to show that $CP$-invariant choices for $\theta_f$ and $\theta_B$ exist iff $\gcd(N,N_f)=1$~\cite{qcdcpdw}. In theories with $\gcd(N,N_f)>1$, matching the anomalies in the infrared provides a fundamental reason why $CP$ must be spontaneously broken at $\theta=\pi$.

These arguments carry over directly to softly broken SQCD, with one small complication. In  QCD, we have only one invariant vacuum angle, $\bar\theta=\theta+\arg\det m$. The $CP$ anomaly is associated with $\bar\theta=\pi$. In SQCD, we have two angles if both $m_\lambda$ and $B$ are nonzero. One combination of them appears as a phase associated with field configurations of nonzero topological charge, and this phase must be $\pi$ in order for the classical $CP$ symmetry to have an 't Hooft anomaly. 

From our analysis of spontaneous $CP$ breaking, we can infer that the relevant phase in softly broken SQCD is $\alpha=\pi$, where $\alpha$ is given in Eq.~(\ref{eq:alpha}). Alternatively, we can consider the vacuum angle dependence of configurations carrying topological charge. The $2N$ gaugino zero modes in an instanton background can be tied off either by bare $m_\lambda$ insertions, or by a self-energy diagram with a quark-squark loop and a $B$-term insertion. Allowing both possibilities for each zero mode, we expect that the instanton amplitude is schematically proportional to\footnote{An extra relative factor of $1/2$ between the two terms appears because our normalization for the gaugino mass in Eq.~(\ref{eq:Vsoft}) is not canonical.}
\begin{align}
{\cal I}\propto \left(m_\lambda-N_f B/32\pi^2\right)^Ne^{i\theta}\;.
\label{eq:I}
\end{align}
Eq.~\ref{eq:I} is rather na\"ive because it uses the loop result in the perturbative vacuum rather than the instanton background.  But $\arg {\cal I}$ then coincides with $\alpha$ of Eq.~(\ref{eq:alpha}), consistent with anomaly matching by spontaneous breaking of CP. This provides evidence that it is the radiatively corrected gaugino mass whose argument controls both effects.

When $CP$ is spontaneously broken but the $CP$ anomaly vanishes, the low energy excitations on the domain walls are not constrained by anomaly inflow, and the wall theories can be trivial in principle. This does not happen in ordinary QCD, where the walls always support Goldstone modes of broken flavor symmetries. However, in SQCD, the $\etap$ is light. We have seen that a necessary condition for the existence of trivial walls is $\gcd(N,N_f)=1$. This coincides with the condition for the vanishing of the 't Hooft anomaly found in~\cite{qcdcpdw}. We have seen that trivial walls may either be approximately BPS, or depend sensitively on SUSY breaking.

When the anomaly  does not vanish, as discussed above, inflow forbids the presence of a trivial wall. We have seen that this is explained in a simple geometric way in SQCD: the vacua are not on branches that are connected in the $\etap$ direction in these theories.

\section{Adjoint QCD with Fundamentals}
\label{sec:qcdadj}

We close by  discussing domain walls in QCD with $N_f$ fundamental fermions $q_f,\bar q_f$, one adjoint fermion $\lambda$, and an axion $\theta$. This theory may also be thought of as SQCD+axion with  squarks and saxion decoupled.\footnote{Implications of spontaneous $CP$ breaking in ordinary QCD on axion physics was recently discussed in~\cite{DiVecchia:2017xpu}.}

Again we can focus on the $\etap$ and $\pi_{-1}$. For small universal quark mass $m_q$ and small adjoint mass $m_\lambda$, the potential is
\begin{align}
V(\etap,\pi_{-1},\theta)\simeq -m_\lambda &\Lambda^3 \cos\left(\frac{\theta+2\pi k+N_f\etap}{N}\right)\nonumber\\
&-m_q \Lambda^3 \left[(N_f-1)\cos(\etap+\pi_{-1})+\cos(\etap-(N_f-1)\pi_{-1})\right]\;,
\end{align}
where $k$ is a branch label $k=0\dots N-1$ for the $\lambda\lambda$ condensate. 

The $\etap,~\pi_{-1},~\theta$, and $k$ degrees of freedom are again subject to various periodicities, or discrete gauge symmetries (DGS). We have the DGS
\begin{align}
&\pi_{-1}\rightarrow\pi_{-1}+2\pi\label{eq:DGSa}\\
&k\rightarrow k+N\label{eq:DGSb}\\
&\etap\rightarrow\etap+2\pi,~k\rightarrow k- N_f \label{eq:DGSc}\\
&\pi_{-1}\rightarrow\pi_{-1} +2\pi/N_f,~\etap\rightarrow\etap-2\pi/N_f,~k\rightarrow k+1\label{eq:DGSd}\\
&\theta\rightarrow \theta+2\pi,~k\rightarrow k-1\label{eq:DGSe}
\end{align}
and combinations thereof. These symmetries can be used to identify different paths in field space that connect vacua.\footnote{Branches and periodicities in QCD with multiple adjoint flavors were discussed in~\cite{Dolan:2017vmn}.}

First, consider the theory with nondynamical axion and $\Lambda\gg m_q\gg m_\lambda$. Here there are  $N$ quasi-degenerate vacua labeled by $k$ (with $\etap=\pi_{-1}=0$) and split by $m_\lambda$.

For $m_\lambda=0$, the vacua are connected by domain walls that traverse pseudo-Goldstone directions in field space. DGS~(\ref{eq:DGSd}) indicates that one trajectory moves in the $\etap$ and $\pi_{-1}$ directions simultaneously. Define 
\begin{align}
\phi\equiv \etap-\pi_{-1}\;.
\end{align}
Holding other directions fixed, the Lagrangian in the $\phi$ direction is
\begin{align}
L(\phi)=\frac{f_{\pi}^2}{8}N_f^2(\partial \phi)^2+m_q\Lambda^3\left[(N_f-1)+\cos\left(\frac{N_f}{2}\phi\right)\right].
\end{align}
The domain wall tension is then of order
\begin{align}
T_\phi\sim f_{\pi}\sqrt{\Lambda^3m_q}\;,
\end{align}
and the profile is of the form
\begin{align}
\phi(z)&=\frac{8}{N_f}\left(\tan^{-1}\left(e^{m_\phi z}\right)\right)\nonumber\\
m_\phi&\equiv \sqrt{m_q\Lambda^3/f_{\pi}^2}\;.
\end{align}

There can also be metastable domain walls in other directions depending on $\gcd(N,N_f)$. From DGS~(\ref{eq:DGSc}), the motion $\etap\rightarrow \etap+2\pi$ connects branch $k$ to branch $k+N_f$, so if $N_f=N-1$, for example, all of the $k$-vacua are traversed in the $\etap$ direction. In particular, for zero $m_\lambda$ each vacuum could be connected by a metastable wall in the $\etap$ direction. The Lagrangian is
\begin{align}
L(\etap)=\frac{f_{\pi}^2}{2}N_f(\partial \etap)^2+m_q\Lambda^3N_f\cos\left(\etap\right)].
\end{align}
Then the tension is of order
\begin{align}
T_{\etap}\sim f_{\pi}N_f\sqrt{\Lambda^3m_q}
\end{align}
and the profile is
\begin{align}
\etap(z)&=4\tan^{-1}\left(e^{m_{\etap} z}\right)\nonumber\\
m_{\etap}&\equiv \sqrt{m_q\Lambda^3/f_{\pi}^2}\;.
\end{align}

For nonzero $m_\lambda$, in general there is only one ground state and none of the domain walls joining the $k$ quasi-vacua are static. However, for fixed $\theta=\pi$, two vacua are degenerate (the $k=0$ and $k=N-1$ vacua for small $m_\lambda$). DGS~(\ref{eq:DGSd}) again indicates that  the $\etap$ and $\pi_{-1}$ directions support a domain wall. 
Likewise there should be a domain wall connecting the degenerate CP-vacua in the $\etap$ direction; this wall may be metastable against cosmic string nucleation.

Now turn on dynamical $\theta$. If $m_\lambda=0$, $\theta$ is a flat direction, and this flat direction smoothly connects the previous $k$-vacua.  If, however,  we give the axion a small mass from another source,
\begin{align}
\Delta V = \tilde\Lambda^4\cos(\theta),
\end{align}
the $N$ vacua remain degenerate. Then we can control whether the stable walls go in the axion or pion directions by adjusting $m_{\pi}/m_a$, where $a=f\theta$ and $m_a\sim \tilde\Lambda^2/f_a$. The axion wall tension is of order
\begin{align}
T_a\sim  f_a \tilde\Lambda^2.
\end{align}

For the opposite hierarchy  $m_\lambda\gg m_q$, and with fixed $\theta$, the $\etap$ has $N_f$ vacua, 
\begin{align}
\etap=-\frac{2\pi k+\theta}{N_f}.
\end{align}
Here only $N_f$ of the $N$ branches give a new vacuum; this is a result of DGS~(\ref{eq:DGSc}). The vacua are degenerate for $m_q=0$, and they are also connected by $\pi_{-1}\rightarrow\pi_{-1}+2\pi/N_f$, which is a flat direction for $m_q=0$. 
Thus  there are metastable static domain walls in the $\etap$ direction for $m_q=0$ and fixed $\theta$. The instability corresponds to a cosmic string composed to the $\etap$ and $\pi_{-1}$ directions. 

For nonzero $m_q$ and fixed $\theta=\pi$, there are again a pair of $CP$-breaking degenerate vacua. As above a stable domain wall is expected in the $\etap$ and $\pi_{-1}$ directions.

With $m_q=0$ and dynamical $\theta$, the axion is another flat direction that connects the vacua under $\theta\rightarrow\theta+2\pi$. There should be an axisymmetric cosmic string solution composed of the axion and $\pi_{-1}$ fields. With small $m_q$, the vacua are split and the two flat directions lifted.

\vskip 1cm
\noindent
{\bf Acknowledgements:}  I thank Adam Nahum, Mithat \"Unsal, and Tin Sulejmanpasic for useful conversations, and especially  Zohar Komargodski for discussions and explanations of $CP$ anomalies and Michael Dine and Laurel Stephenson-Haskins for extensive discussions of BPS domain walls. This work was supported by NSF grant PHY-1719642.

\section*{Appendix}
Here we briefly summarize some of the properties of domain walls in SQCD. The BPS equations have the form
\begin{align}
\partial_z \bar Q_{ff^\prime} + \frac{\partial W^*}{\bar Q_{f f^\prime}^*}e^{-i\delta}=0
\end{align}
and similarly for $Q$. For universal quark masses, the vacua are given in Eq.~(\ref{eq:vvac}),~(\ref{eq:nvac}). $\delta$ may be computed from the constant of motion $E={\rm Im}\left(e^{i\delta}W\right)$, which for a domain wall connecting vacua $n_{1,2}$ reduces to
\begin{align}
E &= Nm v^2 \cos\left(\frac{\pi(n_1-n_2)}{N}\right)\;,\nonumber\\
\delta&=\frac{\pi}{2}-\frac{\pi(n_1+n_2)+\theta}{N}\;.
\end{align}
The tension is 
\begin{align}
T={\rm Re}\left(e^{i\delta}\Delta W\right) = 2N m v^2\sin\left(\frac{\pi(n_2-n_1)}{N}\right)\;.
\end{align}

Some of the domain walls studied in the main text are non-BPS, and furthermore exist only in the presence of soft masses. In this case we must use the  second order Euler-Lagrange equations including the soft potential. 

To construct numerical solutions, we take an ansatz for the $\etap$ and $\pi_{-1}$ trajectories based on the cartoon plots given in the text. Starting from the $\phi_{1,2}$ midpoints, we eliminate another initial condition (in the BPS case) using the conserved quantity, and then use a shooting method to determine any remaining initial conditions.

\bibliography{domain_walls}{}
\bibliographystyle{utphys}

\end{document}